Biological Sciences: Immunology.

# Functional immunomics: Microarray analysis of IgG autoantibody repertoires predicts the future response of NOD mice to an inducer of accelerated diabetes.

[1]Francisco J. Quintana, [2,3]Peter H. Hagedorn, [2]Gad Elizur, [1]Yifat Merbl, [2]Eytan Domany and [1]Irun R. Cohen

The Departments of Immunology[1] and Physics of Complex Systems[2], The Weizmann Institute of Science, Rehovot 76100, Israel. [3]Plant Research Department, Risø National Laboratory, DK-4000 Roskilde, Denmark.

Correspondence to:

Prof. Irun R. Cohen

Department of Immunology, The Weizmann Institute of Science.

Rehovot, 76100, Israel.

E-mail: irun.cohen@weizmann.ac.il

Tel 00972-08934-2911. Fax 00972-08934-4103.

Text pages: 26

Figures: 5

Tables: 1

Words in the abstract: 248

Total number of characters: 50,353

## Abstract

One's present repertoire of antibodies encodes the history of one's past immunological experience. Can the present autoantibody repertoire be consulted to predict resistance or susceptibility to the future development of an autoimmune disease? Here we developed an antigen microarray chip and used bioinformatic analysis to study a model of type 1 diabetes developing in non-obese diabetic (NOD) male mice in which the disease was accelerated and synchronized by exposing the mice to cyclophosphamide at 4-weeks of age. We obtained sera from 19 individual mice, treated the mice to induce cyclophosphamide-accelerated diabetes (CAD), and found, as expected, that 9 mice became severely diabetic while 10 mice permanently resisted diabetes. We again obtained serum from each mouse after CAD induction. We then analyzed, using rank-order and superparamagnetic clustering, the patterns of antibodies in individual mice to 266 different antigens spotted on the chip. A selected panel of 27 different antigens (10% of the array) revealed a pattern of IgG antibody reactivity in the pre-CAD sera that discriminated between the mice resistant or susceptible to CAD with 100% sensitivity and 82% specificity ($p=0.017$). Surprisingly, the set of IgG antibodies that was informative before CAD induction did not separate the resistant and susceptible groups after the onset of CAD; new antigens became critical for post-CAD repertoire discrimination. Thus, at least for a model disease, present antibody repertoires can predict future disease; predictive and diagnostic repertoires can differ; and decisive information about immune system behavior can be mined by bio-informatic technology. Repertoires matter.

## Introduction

Autoimmune diseases are marked by abundant autoantibodies and by vigorously responding T cells targeted to selected self-antigens (1). Immunology has tended to focus on such blatant reactivities (2) and has paid relatively less attention to the autoimmunity detectable to non-classical self-antigens and to the low levels of global autoreactivity detected in healthy subjects (3-6). An important question is whether bioinformatic analysis of the global autoantibody repertoire can predict whether a subject will resist or develop an autoimmune disease before the disease is actually induced by an environmental insult. Can the analysis of immune repertoires contribute to predictive medicine? The present study uses microarray technology and bioinformatic analysis to address that question in an animal model of type 1 diabetes.

Male mice of the NOD strain spontaneously develop type 1 diabetes at a relatively low incidence and late age compared to female NOD mice (7). In our colony, 80-90% of female mice become diabetic by the age of about 6 months compared to about 40-50% of male mice at 9 months of age (8). However, the onset of diabetes can be significantly accelerated and synchronized by exposing NOD mice to cyclophosphamide (9). Cyclophosphamide-accelerated diabetes (CAD) is thought to occur through the selective toxicity of cyclophosphamide for regulatory T cells that otherwise inhibit the disease process (8, 9). The CAD model of type 1 diabetes thus provides an opportunity to test whether the global autoantibody repertoire might reflect resistance or susceptibility to CAD in still-healthy mice, before the cyclophosphamide insult is administered.

We obtained sera from male NOD mice at the age of one month, well before the onset of the spontaneous autoimmune reaction that, once initiated, destroys the β-cells. We then treated the mice with cyclophosphamide and obtained a second sample, after those mice susceptible to diabetes had developed the disease. In this way, we could test the pre-CAD and the post-CAD sera from both susceptible and resistant mice. Recently, microarray antigen chips have been used to detect high-titer autoantibodies to known antigens in autoimmune diseases (10, 11). However, rather than focusing only on known self-antigens, we here profiled individual immune systems by their global patterns of autoantibodies free of bias for high-titer reactivities to particular self-antigens. We developed a microarray antigen chip (manuscript in preparation) by covalently spotting 266 different antigens to the coated surface of glass slides, incubated these antigen chips with the sera of the individual mice obtained before and after CAD induction and detected by laser illumination the amounts of antibodies binding to the different antigen spots. Since type 1 diabetes is caused by autoimmune T cells (12), we focused the repertoire analysis on the IgG antibodies, whose presence implies T-cell reactivity.

## Materials and Methods

*Mice*

Male NOD mice were raised and maintained under pathogen-free conditions in the Animal Breeding Center of The Weizmann Institute of Science. The experiments were carried out under the supervision and guidelines of the Animal Welfare Committee. The mice were 4-weeks old at the start of the experiments. Nineteen mice were studied individually.

*CAD*

Diabetes onset was accelerated and synchronized as previously described (9) by two intraperitoneal injections of 200 mg/kg of cyclophosphamide (Sigma, Rehovot, Israel) given at the age of 4 weeks and again one week later. In our colony, this treatment of NOD males leads to an incidence of diabetes of about 50 % (8). The mice developing diabetes go on to die unless they are treated with insulin; those males that do not develop diabetes within 1 month after two injections of cyclophosphamide do not become diabetic later in life (data not shown). Figure 1 is a schematic representation of the protocol.

*Diabetes*

Blood glucose was measured weekly. A mouse was considered diabetic when its blood glucose concentration was higher than 13 mM on two consecutive examinations, tested using a Beckman Glucose Analyzer II (Beckman Instruments, Brea, California, USA). Of the 19 mice treated with cyclophosphamide, 9 developed diabetes and 10 remained healthy throughout a 2-month period of observation.

*Sera*

Serum samples were collected one day before the first injection of cyclophosphamide, and 1 month after the second injection. Blood was taken from the lateral tail vein, allowed to clot at room temperature and, after centrifugation, the sera were stored at - 20°C.

*Antigens*

The 266 antigens spotted on the microarray chips in these studies include proteins, synthetic peptides from the sequences of key proteins, nucleotides and phospholipids, and are enumerated in the Supporting Information, Table 1.

*Antigen microarray chips*

Antigens diluted in PBS were placed in 384-well plates at a concentration of 1 μg/μl. We used a robotic MicroGrid arrayer with solid spotting pins of 0.2 mm diameter (BioRobotics, Cambridge, UK) to spot the antigens onto ArrayIt SuperEpoxi Microarray Substrate slides (TeleChem International, Sunnyvale, California, USA). Each antigen was spotted in 2-8 replicates. The spotted microarrays were stored at 4°C. The chips were washed with PBS and blocked for 1 hr at 37°C with 1 % BSA, and incubated overnight at 4°C with a 1/5 dilution of the test serum in blocking buffer under a cover slip. The arrays were then washed and incubated for 45 minutes at 37°C with a 1/500 dilution of a goat anti-mouse IgG Cy3-conjugated antibody, purchased from Jackson ImmunoResearch Labs. Inc. (West Grove, Pennsylvania, USA). The arrays were washed again, spun dried and scanned with a ScanArray 4000X scanner (GSI Luminomics, Billerica, Massachusetts, USA). The results were recorded as TIFF files.

*Image and data processing*

The pixels that comprised each spot in the TIFF files and the local background were identified using histogram segmentation. The intensity of each spot and its local background were calculated as the mean of the corresponding pixel intensities. None of the spots containing antigens showed saturation. Technically faulty spots were identified by visual inspection and removed from the dataset. For each spot, the local background intensity was subtracted from the spot intensity. Spots with negative intensities were removed from the dataset. A log-base-2 transformation of the intensities resulted in reasonably constant variability at all intensity levels. The log-intensity of each antigen was calculated as the mean of the log-intensities of the replicates on each slide. The coefficient of variability (CV) between replicates on each array was around 30%.

To remove overall differences in intensities between arrays, the mean-log-intensity of each antigen on each array was scaled by subtracting the median of the mean-log-intensities of all antigens on the array. The scaled mean-log-intensity of an antigen is denoted the *reactivity of the antigen*.

The processed dataset consists of a matrix of IgG reactivities consisting of 266 rows and 38 columns (2 samples for each of 19 mice). Each column contains the reactivities measured on a given array and each row contains the reactivities measured for a given antigen over all arrays.

Additionally, the reactivity for each antigen measured before and after cyclophosphamide treatment in each mouse was combined into a log-ratio by subtracting the reactivity before treatment from the reactivity after treatment. This yielded a matrix of ratios with 266 rows and 19 columns.

*Data analysis*

We based the clustering of antigens and samples on the Superparamagnetic Clustering (SPC) algorithm (13) because it provides an inherent mechanism for identifying robust and stable clusters. The algorithm can be understood by an analogy to physics: as a parameter $T$ (the temperature) is increased, the system undergoes phase transitions (for example, it melts). In our case, $T$ is increased from 0 (all objects form one cluster) to $T_{max}$ (each object forms a separate cluster). The break up of larger clusters into smaller sub-clusters is governed by the structure of the data: similar objects tend to stay together over a large increase in $T$, while less similar objects break apart more easily. The range of $T$'s for which a given cluster remains unchanged, denoted by $\Delta T$, is used as a stability measure for the cluster. As the measure of similarity between objects, we used Euclidean distance for both samples and antigens. Since the antigen reactivities (or ratios) were first row-centered and normalized before being clustered, their squared distance is proportional to $1 - r$, where $r$ is the correlation coefficient. The correlation coefficient captures similarity in shape and the Euclidean distance captures similarity in magnitude.

To determine subsets of the 266 antigens that would separate the sick and healthy mice, we used the Wilcoxon rank-sum test (14). This test is non-parametric; it is robust to outliers. We test one antigen at a time, replacing the reactivities (or ratios) with ranks according to their magnitude: 1 for the smallest, 2 for the second smallest, and so on. The $p$-values found using this method were higher than 0.01; but no single antigen was found to significantly discriminate between the two groups when the Bonferroni-correction (15) or the False Discovery Rate method (16) were applied. This means that the signal produced by any single antigen is unable to separate the sick mice from the healthy. Separability might be achieved, if at all, by using reactivity (or ratio) profiles defined over several antigens. To capture a collective effect of several antigens, we selected the 27 antigens (10% of the 266 antigens in the study) with the lowest $p$-values, and investigated how good they were collectively at separating sick from healthy mice, and which antigens showed correlated behavior over the samples, by applying two-way SPC. This gives an unsupervised clustering of the subset of antigens and of the samples. The clusters of samples found using this method were evaluated for their stability $\Delta T$, specificity, and sensitivity. Specificity is the proportion of sick mice in the "sick cluster"; sensitivity is the proportion of the sick mice in the "sick cluster" compared to all the sick mice in the study.

*Statistical significance*

To obtain a measure of the significance of the separation between sick and healthy mice using the method outlined above, we performed the following test: From the group of healthy mice, we picked at random 5 of the samples, and similarly, for the group of sick mice we randomly picked 4 of the samples. These 9 samples were labeled as "type A". The remaining samples were labeled as "type B". We hypothesized that there should be no clear separation between these randomized types: we used the Wilcoxon rank-sum test to identify the 27 antigens that differentiated best groups A and B. Next, we clustered the mice in the space of these 27 antigens, and look for stable, specific, and sensitive clusters, using SPC. We performed the test 1000 times on different randomized groups and recorded the stability, specificity, and sensitivity of the resulting clusters. The proportion of random clusters manifesting these features to the same or to a better degree than the actual cluster establishes the *p*-value of the actual cluster.

## Results

*Selection of informative antigens for pre-CAD mice*

We have previously reported that coupled two-way clustering (CTWC) could be used to successfully separate human subjects already diabetic from healthy persons (17). In the CAD mouse study done here, however, only a few of the clusters of co-regulated antigens using the CTWC technique separated between the sick and the healthy mice before CAD, and then only for a subset of the mice. We therefore took a different approach. Based on the sera taken before cyclophosphamide treatment, Table 1 List I tabulates the 27 antigens that separated best between the sera of the 10 mice that later resisted the induction of CAD and the 9 mice that later developed CAD (using the Wilcoxon rank-sum test and taking the 10% with lowest *p*-values).

Figure 2 Left Panel, shows the two-way SPC of these antigens. The mice susceptible to future CAD induction are denoted by the filled rectangles at the top of the clustering box; the mice resistant to future CAD induction are denoted by the empty rectangles. The 27 antigens are clustered at the rows, and identified by number (see Table 1). It can be seen that all 9 mice that were found later to be susceptible to CAD could be separated from 8 of the 10 mice that were later found to resist CAD; before cyclophosphamide, the CAD-susceptible mice manifested relatively elevated IgG reactivity to the top 19 antigens in Figure 2 Left Panel, while the CAD-resistant mice manifested relatively elevated IgG reactivity to the remaining 8 antigens. The clustering separation was significant ($p<0.017$; only 17 of 1000 randomly generated groups showed results comparable to the actual data set). Thus mice susceptible to CAD could be distinguished by their patterns of IgG serum antibodies from mice resistant to CAD, even before cyclophosphamide was administered to the mice.

*Selection of informative antigens for post-CAD mice*

We then used the 27 antigens effective in pre-CAD clustering to analyze the patterns of IgG antibodies developing in the diabetic and healthy mice post-CAD. Surprisingly, these 27 antigens failed to discriminate between the two groups of mice; the obvious pre-CAD clusters seen in Figure 2, Left Panel, dispersed when the same antigens were used to cluster the post-CAD sera; compare the Right and Left panels in Figure 2. For this reason, we tested whether other sets of antigens might be more informative post-CAD. The 27 antigens listed in Table 1 List II was generated by performing the Wilcoxon rank-sum test on the reactivities measured post-CAD. A third set of 27 antigens was generated by performing the Wilcoxon rank-sum test on the ratios by which each antigen changed post-CAD/pre-CAD. The ratios provide information on reactivity changes toward the antigen. These antigens are shown in Table 1 List III.

Figures 3 and 4 show that the List II and the List III antigens could indeed separate between the healthy and diabetic mice post-CAD: specificity up to 82% and sensitivity up to 100% ($p=0.065$). Thus, the IgG repertoires of the pre-CAD and post-CAD groups of healthy and sick mice could be clustered, but the informative patterns of reactivity required modified sets of antigens to develop discriminating patterns.

It can be seen that some of the antigens from the set of pre-CAD antigens (Table 1 List I) were also present in the post-CAD set (Table 1 List II), or in the set of antigens determined from the pre-CAD/post-CAD ratios (Table I List III). For example, three of the pre-CAD antigen reactivities were also prominent post-CAD (antigens 17, 18 and 26; Table 1 List I). The shared and distinct antigens are shown as a Venn diagram for the overlap between Lists I, II and III in Figure 5. List III in Table 1 (ratio difference) can be seen to have generated a set of antigens most shared (dark rectangles) between pre-CAD sera (List I) and post-CAD sera (List II); see Table 1 and Figure 5.

## Discussion

The antigen microarray chip described in this paper required much preliminary work to obtain consistent results, including determination of a workable surface coating for the glass, reagent concentrations and incubation times, size of spots, distances between spots, washing protocols, laser activation and reading, and other technical issues (manuscript in preparation). Patterns of IgM antibodies were analyzed both before and after CAD, but these results too will be presented elsewhere.

Here we show that the patterns of IgG antibodies expressed pre-CAD in male NOD mice can mark susceptibility or resistance to CAD induced later. We also found patterns of IgG antibodies characteristic of healthy or diabetic mice post-CAD, but these patterns required sets of antigens that differed from the informative pre-CAD set (see Table 1). Thus, IgG reactivities to some antigens may mark future susceptibility to CAD, but not CAD itself once the disease emerges, and, conversely, some IgG reactivities may mark the disease but not the susceptibility. Hence, prediction of future disease (this paper) and diagnosis of present disease (this paper and (10, 11)) can depend on different data sets of information, at least in the CAD model. The reasons for this divergence need to be investigated, but the divergence itself may be explained by the likelihood that the IgG antibodies we measured are not themselves the causal agents, but only indirect, surrogate markers for the autoimmune T cells that directly regulate or mediate the diabetic process. This observation should alert us to the possibility of a similar divergence between the prediction and the diagnosis of human diseases.

Another notable finding was that health, both pre-CAD and post-CAD, was associated with relatively high IgG autoreactivity to self-antigens, to which the susceptible mice were low IgG responders (Figures 2 and 4, and Table 1). Thus some types of active autoimmunity may actually protect against autoimmune disease (18-20). This finding is compatible with the idea that autoimmunity of certain specificities is not only compatible with health, but essential for health (21, 22).

Individual mice of the highly inbred NOD strain would seem to bear very similar, if not identical genomic DNA, yet almost half the male mice resist CAD as well as they resist slowly progressive spontaneous diabetes. In humans, too, type 1 diabetes develops in persons bearing certain alleles, predominantly alleles of HLA immune response genes (23), but most individuals who have inherited these susceptibility alleles will never develop the disease. Indeed, identical twins develop type 1 diabetes with a concordance rate of less than 50%, despite having inherited identical genomic DNA (24). Thus, environmental factors would appear to determine whether the diabetic potential inherent in one's genome becomes realized as type 1 diabetes (25, 26). Since type 1 diabetes, including the CAD variant in NOD mice, is an autoimmune disease (7, 9, 12), it is very likely that resistance or susceptibility to the disease emerges from the interaction of the individual's immune system with the environment, down-stream of a permissive germ-line genetic endowment. Indeed, only the changing environment can be blamed for the alarming increase in the incidence of type 1 diabetes noted in the past few decades; significant changes in the frequencies of human genes have not occurred in the interim in affected populations (27). Thus, type 1 diabetes emerges from the impact of the environment on the structure and function of the immune system in a way that transforms naturally benign autoimmunity into an autoimmune disease affecting the insulin-producing β cells (22). Various environmental factors can probably act to induce type 1 diabetes in susceptible individuals; the CAD model is one example.

But environmental factors can also prevent the development of type 1 diabetes. Stimulation of the NOD mouse immune system by infection (28-30) or by vaccination with microbial antigens (31) or by treatment with ligands that activate innate immune receptors (32-34) can prevent diabetes. Thus, the cumulative experience of the immune system (including, for example, positive autoimmunity to antigens such as the lower 8 antigens in Figure 2, Left Panel) can determine the organization of its component molecules and cells regarding self-antigens, and this internal structuring can, in turn, help one resist the accidental induction of an autoimmune disease. Type 1 diabetes appears in very young people (12, 35), so critical aspects of autoimmune organization must occur fairly early in one's lifetime. The results of this bioinformatic study would suggest that, in addition to individual differences in immune repertoires, some organized patterns of IgG autoantibodies are shared by groups of individuals, at least among NOD mice.

The bioinformatic analysis described here relates to two separate, but linked issues: predictive medicine via functional immunomics and the biological meaning of the autoimmune repertoire.

*Functional immunomics* may be defined as the functional state of the immune system inscribed in its global patterns of immune molecules and cells. Functional immunomics, even that limited to part of the IgG autoantibody repertoire as we show here, can help anticipate disease before it emerges, and anticipation is an important first step in predictive medicine.

Beyond its potential usefulness for predictive medicine, functional immunomics may teach us some things about the biology of immune system organization. Note that the list of informative antigens (Table 1) does not contain insulin, a well-studied self-antigen in diabetes (12). A peptide of glutamic acid decarboxylase (GAD), used clinically to diagnose type 1 diabetes in humans (2, 36, 37), was only informative after CAD was induced (Table 1 List II). The immune system is a complex system, and reactivities of seemingly minor magnitude can play major roles in complex system behavior (22, 38). Measuring autoantibodies to a few known antigens only (10, 11) may not provide the same information as can a global pattern.

The present study investigated patterns of antibodies, and not their function in the disease process. Nevertheless, the list of informative antigens may be connected to other observations regarding the pathophysiology type 1 diabetes. Six of the 8 antigens to which relatively high IgG reactivity is associated with resistance to CAD are peptides derived from heat shock proteins (HSP): peptides p277, 22 and 16 of HSP60, peptides 1 and 7 from the sequence of GroEl (the HSP60 molecule of *E. coli*), and peptide 13 of HSP70. Indeed, the three antigens associated with health both before and after CAD are p277, peptide 22 of HSP60 and peptide 1 of GroEl (see Figure 5 and Table 1). Vaccination with HSP60/p277 can arrest type 1 diabetes in NOD mice (18, 39), and has been shown to arrest the destruction of insulin-producing beta cells in a clinical trial in humans (40).

It is worthy of note that peptide p277 of HSP60 was first discovered as a dominant epitope for T cells (18). The natural IgG reactivity of CAD-resistant mice to this "T-cell peptide" demonstrates that some autoantibodies do reflect elements of the T-cell repertoire. Indeed, prevention of spontaneous diabetes in NOD mice by stimulating innate toll-like receptors with CpG oligonucleotide was found to spontaneously activate the production of IgG antibodies to peptide p277 (34). We have found that vaccination with HSP60 can inhibit CAD, apparently by modifying the cytokine profile of autoimmune effector T cells (19). HSP60 vaccination can also induce regulatory T cells effective in models of autoimmune arthritis (20, 41). Thus the association of resistance to CAD with natural IgG antibodies to HSP60 peptides suggests that medicinal vaccination with HSP60 or its peptides may work by strengthening regulatory networks that arise naturally through immune experience with endogenous (or cross-reactive bacterial) heat shock proteins. Autoimmunity to HSP60 peptides like p277 is built into the healthy immune system (22).

The 19 antigens targeted by IgG antibodies in the CAD-susceptible mice are also interesting biologically. Three peptides of HSP70 are included, and T-cell autoimmunity to HSP70 has been described in human type 1 diabetes patients (42). Gliadin is an antigen associated with Celiac disease, and celiac patients have been reported to have an increased incidence of type 1 diabetes (43). MOG is a molecule present in myelin, and T-cell autoimmunity to MOG can induce experimental autoimmune encephalomyelitis in NOD mice (44). Glucagon is produced by α cells in the pancreatic islets adjacent to the β cells that produce insulin, but no studies of autoimmunity glucagon have been yet reported in type 1 diabetes. Accelerated atherosclerosis is a serious complication of type 1 diabetes (45), and is assumed to arise as a complication of poor glucose homeostasis in poorly controlled diabetes (46). Autoimmunity to LDL and HDL, however, has been proposed to be a factor in atherosclerosis in general (45). The finding of heightened IgG autoimmunity to LDL and HDL in the mice susceptible to CAD suggests the possibility that LDL and HDL autoimmunity might actually be part of the collective of autoimmune reactions responsible for the primary development of type 1 diabetes. If this is true, then the vascular "complications" of type 1 diabetes may be a primary and early event in the disease process and not merely a phenomenon secondary to poor metabolic control. VEGF and vasopressin too are molecules that function in blood vessel formation and the physiology of blood flow (47, 48). The increase in IgG antibodies to certain antigens post-CAD is also intriguing, and we will discuss the possible biological significance of these reactivities elsewhere. For now, it is important to note that a bioinformatic analysis can, by itself, raise new questions for further biological research; arrays of antigens open new windows for viewing natural autoimmunity, autoimmune disease and the links between them.

The demonstration of patterns of autoantibody reactive with key self-molecules and the association of such reactivity with health challenges basic assumptions of the classical clonal selection theory (CST) of adaptive immunity (49). According to the CST, autoimmune repertoires should not exist in healthy individuals. The present findings are more compatible with a cognitive paradigm of immunity (22, 50, 51).

The core of organized autoimmune repertoires within the immune system has been termed the immunological homunculus, the immune system's internal representation of the body under its care (50, 51). The mammalian immune system, in addition to its well-studied role in defending the body against foreign invaders, is now understood to be heavily involved in maintaining the integrity of the body from within; immune system cells and molecules, which comprise the inflammatory response, are key factors in wound healing, neuroprotection, connective tissue formation, angiogenesis, tissue morphology and regeneration, and waste disposal (5, 21, 22). To dispense reparative inflammation at the right sites and occasions, the immune system has to assess the state of the body on the fly. In this respect, the immune system acts as it were the body's onboard bioinformatic computer. If so, predictive medicine would do well to mine this immune information, as this study suggests it might.

## Acknowledgements

IRC is the incumbent of the Mauerberger Chair in Immunology. ED is the incumbent of the H J Leir Professorial Chair. The project was partly funded by the Chief Scientist's Office of the Israeli Ministry of Industry and Trade, and by Compugen. PHH was supported by a fellowship of the European Community.

## Figure Legends

**Figure 1.** The experimental protocol. The numbers refer to the age (in weeks) of the mice. The black vertical lines at 4 and 9 weeks indicate serum sample collection. The grey vertical lines at 4 and 5 weeks indicate cyclophosphamide injection. The grey box at 6 weeks shows when the CAD-susceptible mice developed diabetes, and the grey box between 11 and 13 weeks shows the time of death of the untreated diabetic mice.

**Figure 2.** Reactivity matrices of 27 antigens separate diabetic and healthy mice, before CAD induction. The rows are antigens and the columns are the mouse sera. Each antigen is identified by the number shown between the two reactivity matrices (see Table 1). The *Left Panel* shows a two-way SPC of the antigens and the serum samples pre-CAD. The length of a branch connecting to a cluster represents the stability of the cluster. Filled boxes denote mice that later developed CAD, and open boxes denote mice that resisted CAD. The *Right Panel* shows the SPC of the serum samples post-CAD. Filled boxes denote mice that developed CAD, and open boxes denote healthy mice that resisted CAD. The antigens used in the two panels are the same and presented in the same order.

**Figure 3.** Two-way SPC of 27 antigens that separate sick and healthy mice, post-CAD. Filled boxes denote diabetic mice and open boxes denote healthy mice.

**Figure 4.** Two-way SPC of 27 antigens that separate the sick and healthy mouse samples using the pre-CAD and post-CAD ratios. Filled boxes denote mice susceptible to CAD and open boxes denote mice resistant to CAD.

**Figure 5.** Venn diagram showing antigens shared by the three lists of 27 antigens: (I), pre-CAD; (II), post-CAD; and (III), sick and healthy mice by ratio. See Table 1.

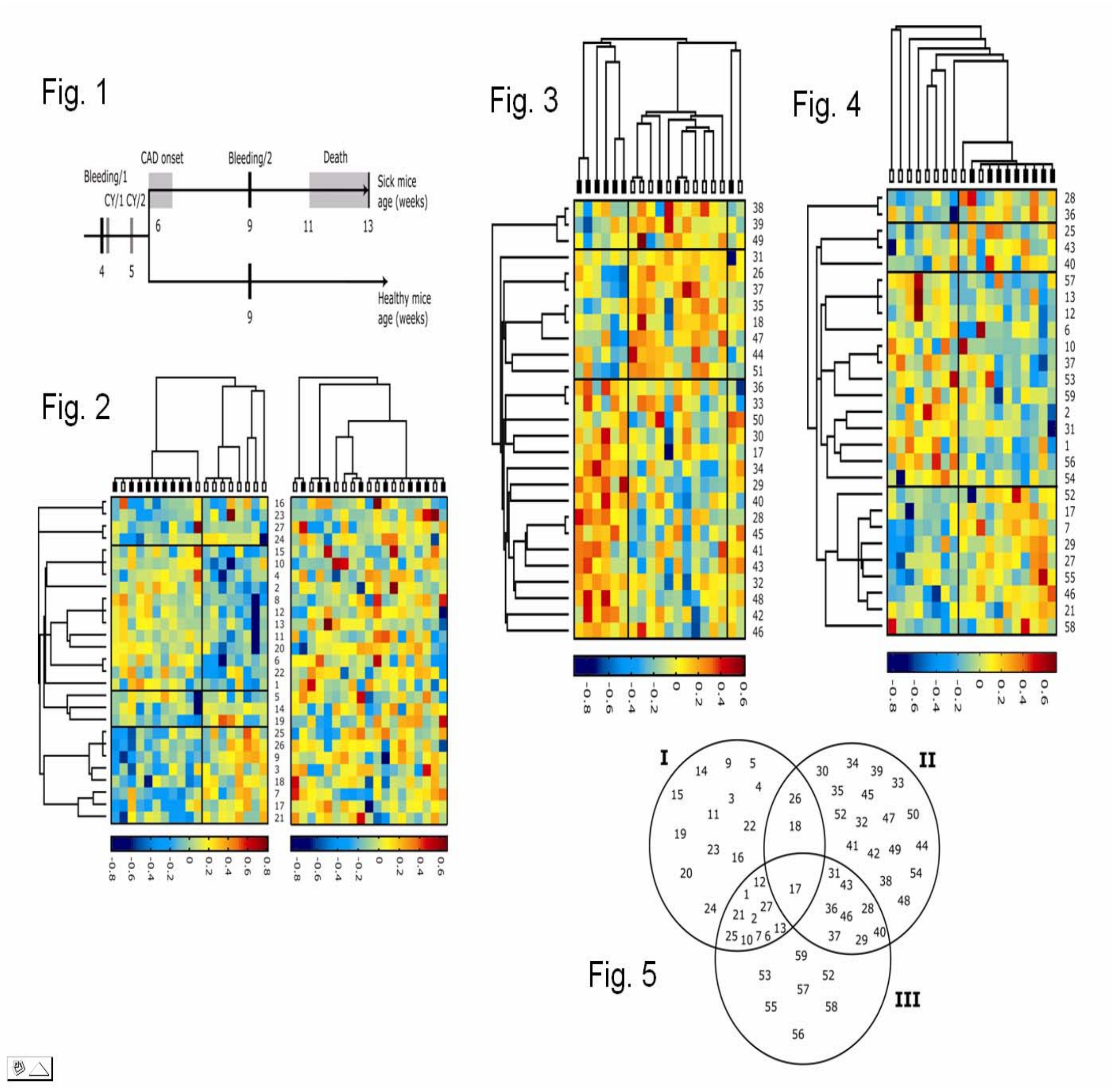

**Table 1. Discriminating sets of antigens.**

| Class | Antigen | # | Antigen Lists I | II | III |
|---|---|---|---|---|---|
| HSP | HSP60/p16 | 21 | H | | S |
| | HSP60/p22 | 17 | H | S | S |
| | HSP60/p30 | 55 | | | S |
| | HSP60/p34 | 36 | | S | S |
| | HSP60/p35 | 59 | | | H |
| | HSP60p277 | 26 | H | H | |
| | GroEL/p1 | 18 | H | H | |
| | GroEL/p7 | 9 | H | | |
| | GroEL/p10 | 35 | | H | |
| | GroEL/p11 | 31 | | H | H |
| | GroEL/p15 | 47 | | H | |
| | GroEL/p16 | 51 | | H | |
| | GroEL/p18 | 44 | | H | |
| | GroEL/p23 | 37 | | H | H |
| | GroEL/p25 | 53 | | | H |
| | GroEL/p28 | 56 | | | H |
| | HSP70/p4 | 48 | | S | |
| | HSP70/p6 | 46 | | S | S |
| | HSP70/p8 | 10 | S | | H |
| | HSP70/p9 | 32 | | S | |
| | HSP70/p13 | 7 | H | | S |
| | **HSP70/p12** | 38 | | **H** | |
| | **HSP70/p17** | 22 | **S** | | |
| | **HSP70/p22** | 28 | | **S** | **S** |
| | **HSP70/p23** | 49 | | **H** | |
| | **HSP70/p24** | 14 | **H** | | |
| | **HSP70/p30** | 15 | **S** | | |
| | **HSP71** | 19 | **H** | | |
| **Tissue antigens** | **Glucagon** | 12 | **S** | | **H** |
| | **GAD/p34** | 29 | | **S** | **S** |
| | **C-peptide** | 41 | | **S** | |
| | **MOBP/p78-89** | 40 | | **S** | **S** |
| | **MOG mouse** | 24 | **H** | | |
| | **Cartilage Extract** | 52 | | | **S** |
| | **Vimentin** | 57 | | | **H** |
| | **VEGF** | 8 | **S** | | |
| **Immune receptors** | **TCR β–chain/pMed12** | 30 | | **S** | |
| | **TCR β–chain/pN12** | 43 | | **S** | **S** |
| | **IL-2R α–chain/p2** | 34 | | **S** | |
| | **IL-2R β–chain/p1** | 45 | | **S** | |
| **Enzymes** | **Acid Phosphatase** | 3 | **H** | | |
| | **Aldolase** | 33 | | **S** | |
| | **Collagenase** | 39 | | **H** | |
| | **GSTase** | 50 | | **H** | |
| | **holo-transferase** | 27 | **H** | | **S** |
| **Hormones** | **βMSH** | 2 | **S** | | **H** |
| | **BNP** | 20 | **S** | | |
| | **DAP** | 11 | **S** | | |
| | **Gliadin** | 23 | **H** | | |
| | **Somatostatin** | 54 | | | **H** |
| | **Vasopresin** | 6 | **S** | | **H** |
| | **VIP** | 13 | **S** | | **H** |
| **Plasma** | **Plasmin** | 42 | | **S** | |
| | **HDL** | 5 | **S** | | |
| | **LDL** | 1 | **S** | | **H** |
| | **human Serum Albumin** | 58 | | | **S** |
| | **methylated BSA** | 4 | **S** | | |
| **Other antigens** | **KLH** | 25 | **H** | | **S** |
| | **PS4** | 16 | **H** | | |

The numbers (#) refer to the antigens shown in the Figures. The dark boxes indicate the antigens that participated in the separations. The letter in the box indicates the group in which the reactivity to the antigen (or pre-CAD/post-CAD ratio) was relatively the highest; **S** designates the sick group of mice, and **H** designates the healthy group of mice. **Antigen List I** refers to the 27 antigens pre-CAD selected by rank-sum from those remaining healthy (H) and those later developing diabetes (S) after cyclophosphamide. **Antigen List II** refers to the 27 antigens selected by rank-sum for the healthy and sick groups post-CAD. **Antigen List III** refers to the 27 antigens selected by rank-sum from the pre-CAD/post-CAD ratios.

Antigen abbreviations: **GAD**, glutamic acid decarboxylase**; MOBP**, Myelin-Associated Oligodendrocytic Basic Protein**; MOG**, Myelin Oligodendrocyte Glycoprotein**; VEGF**, Vascular Endothelial Growth Factor; **TCR**, T-cell receptor; **GSTase,** Galactosyltransferase; **MSH,** Melanocyte Stimulating Hormone; **BNP,** Brain Natriuretic Peptide; **DAB**, Diabetes Associated Peptide amide; **VIP**, Vasointestinal peptide; **HDL**, Lipoprotein, High Density; **LDL**, Lipoprotein, Low Density; **BSA**, Bovine Serum Albumin; **KLH**, Keyhole Lympet Hemocyanin.